\documentclass[prl,twocolumn,superscriptaddress,showpacs,amsmath,amssymb]{revtex4-2}
\usepackage{graphicx, color}
\usepackage{dcolumn}
\usepackage{bm}
\usepackage{epstopdf}
\usepackage{physics}
\usepackage[usenames,dvipsnames]{xcolor}
\usepackage{subfigure}
\usepackage{pifont}

\usepackage{verbatim}

\begin{document}
\title{Reducing the frequency of the  Higgs mode in a helical superconductor \\ coupled to an LC-circuit}

\date{\today}
\begin{abstract}

We show that the amplitude, or Higgs mode of a superconductor with strong spin-orbit coupling and an exchange field, couples linearly to the electromagnetic field. Furthermore, by coupling such a superconductor to an LC resonator, we demonstrate that the Higgs resonance becomes a regular mode at frequencies smaller than the quasiparticle energy threshold $2\Delta$. We finally propose and discuss a possible experiment based on microwave spectroscopy for an unequivocal detection of the Higgs mode. Our approach may allow visualizing Higgs modes also in more complicated multiband superconductors with  a coupling between the charge and other electronic degrees of freedom.

\end{abstract}

\author{Yao Lu}
\email{yao.lu@ehu.eus}
\affiliation{
Centro de F\'{i}sica de Materiales (CFM-MPC), Centro Mixto CSIC-UPV/EHU, Manuel de Lardizabal 5, E-20018 San Sebasti\'{a}n, Spain}

\author{Stefan Ili\'{c}}
\affiliation{
Centro de F\'{i}sica de Materiales (CFM-MPC), Centro Mixto CSIC-UPV/EHU, Manuel de Lardizabal 5, E-20018 San Sebasti\'{a}n, Spain}

\author{Risto Ojaj\"arvi}
\affiliation{Institute for Theory of Condensed Matter, Karlsruhe Institute of Technology (KIT), 76131 Karlsruhe, Germany}

 \author{Tero T. Heikkil\"a}
 \email{tero.t.heikkila@jyu.fi}
\affiliation{Department of Physics and Nanoscience Center, University of Jyväskylä, P.O. Box 35 (YFL), FI-40014 University of Jyvaskyla, Finland}

\author{F. Sebastian Bergeret}
\email{fs.bergeret@csic.es}
\affiliation{
Centro de F\'{i}sica de Materiales (CFM-MPC), Centro Mixto CSIC-UPV/EHU, Manuel de Lardizabal 5, E-20018 San Sebasti\'{a}n, Spain}
\affiliation{Donostia International Physics Center (DIPC), Manuel de Lardizabal 4, E-20018 San Sebasti\'{a}n, Spain}

\maketitle
\paragraph{Introduction.-}

\begin{figure}[t]
\centering
\includegraphics[width = \columnwidth]{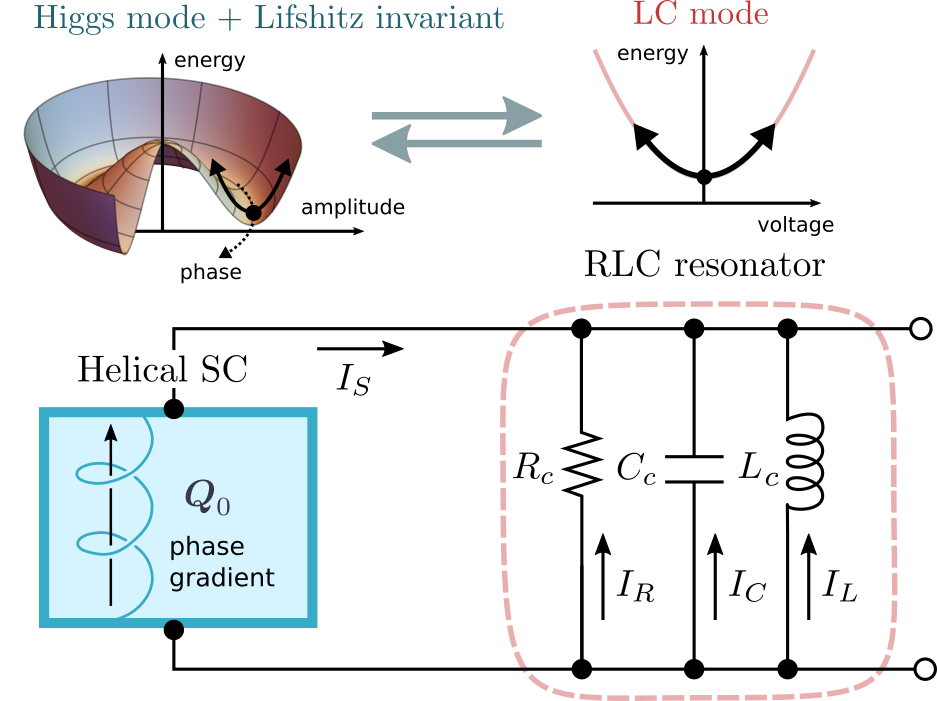}
\caption{Schematic diagram of the circuit that couples a helical superconductor and an LC resonator. Microwave is sent to the system through a transmission line, and the resonance modes are detected by measuring the microwave reflection rate. The Lifshitz invariant couples the supercurrent and the Higgs mode.}\label{Fig:Schematic}
\end{figure}

In superconductors with spontaneously broken $U(1)$ symmetry, the  Higgs mode is an excitation associated with the oscillation of the  order parameter  amplitude around its saddle point value \cite{higgs,Higgsmode1,particleSC,littlewood1981gauge,littlewood1982amplitude}, see Fig.~\ref{Fig:Schematic}. Despite the progress in studying the Higgs mode in systems where charge-density-wave order and superconductivity coexist \cite{higgscdw1,higgscdw2,higgscdw3,higgscdw4,littlewood1982amplitude}, or via its non-linear coupling to the electromagnetic (EM) field \cite{Thzhiggs1, Thzhiggs2, Thzhigg3,mikhail1,mikhail2}, detecting unequivocally the Higgs mode, in general,  remains a challenging task. 

One of the obstacles is that the Higgs mode in conventional superconductors is a scalar mode, and hence it couples nonlinearly to the EM field. A linear coupling can be achieved in the presence of a supercurrent \cite{moor2017amplitude}. 
A second reason for its challenging detection is that  
with a mass of $2\Delta$, the Higgs mode resides precisely at the bottom of the quasiparticle continuum
, and is thus over-damped by the quasiparticle excitations. Unlike a regular collective mode, the Higgs mode corresponds to a square root singularity of the pair susceptibility. As a result, it decays in time in a power law fashion $\delta\Delta(t)\sim\delta\Delta(0)\cos(2\Delta_0t)/\sqrt{2\Delta_0t}$ \cite{volkov1974collisionless}, where $\delta\Delta$ is a small perturbation of the pairing gap around its  average value, $\Delta_0$. Even though it was suggested that the mass of the Higgs mode can be  below the energy gap in strongly disordered superconductors \cite{sherman2015higgs,sacepe2010pseudogap,sacepe2011localization,mondal2011phase,chand2012phase,noat2013unconventional,kamlapure2013emergence,ghosal2001inhomogeneous,bouadim2011single}, it was shown later that in such  systems, the Higgs mode  never shows up as a real mode \cite{cea2015nonrelativistic}.

In this letter, we propose a way  of overcoming these  difficulties. 
 We first demonstrate that the amplitude mode in helical superconductors \cite{agterberg2003novel,kaur2005helical,dimitrova2007theory} couples linearly to the EM field, even in the absence of  a supercurrent. Helical superconductivity occurs in systems where both inversion and time-reversal symmetries are broken, for instance due to magnetic fields and spin-orbit coupling (SOC). Secondly,  we exploit  such a linear coupling and demonstrate the  reduction of  the Higgs frequency  by coupling  the superconductor to an LC resonator (Fig.~\ref{Fig:Schematic}). When the resonant frequency of the LC mode is slightly larger than the Higgs frequency,  and the direct coupling between the two modes is finite, they repel each other, and the Higgs mode is pushed down to frequencies smaller than  $2\Delta$ making it a well-defined mode.

\paragraph{Summary of the results from a phenomenological model.-}In a conventional superconductor the Higgs-light coupling is characterized by the susceptibility
\begin{equation}
\chi_{\boldsymbol{A}\Delta}=\frac{\partial^2 S}{\partial \boldsymbol{A}\partial\Delta}=\frac{\partial\boldsymbol{J}}{\partial\Delta}\;,\label{eq:Higgslightcoupling}
\end{equation}
where $S$ is the action, $\boldsymbol{A}$ is the vector potential, and $\boldsymbol{J}={\partial S}/{\partial \boldsymbol{A}}$ is the supercurrent. Near the critical temperature,  $\boldsymbol{J}\propto\Delta^2$, and  the susceptibility is  $\chi_{\boldsymbol{A}\Delta}\propto {J}/\Delta$.  In other words,  it is finite only in the presence of  a supercurrent  \cite{moor2017amplitude}. 

The situation is different in 
superconductors with broken time-reversal  and inversion symmetries. These may correspond to superconductors with Rashba SOC and an in-plane exchange field,  intensively studied in the context of magnetoelectric phenomena in superconductors, such as helical superconductivity \cite{agterberg2003novel,kaur2005helical,dimitrova2007theory}, Josephson $\phi_0$ junctions \cite{buzdin2008direct,bergeret2015theory}, and most recently supercurrent diode effects \cite{diodeexp,diodethe1,diodethe2,diodethe3,ilic2022theory,he2022phenomenological}.  In this case, the action up to the fourth order in the order parameter is given by  \cite{he2022phenomenological,ilic2022theory},
\begin{equation}
\begin{split}
     S=&S_0+\int dtd\boldsymbol{r} \left(\boldsymbol{a}_1\tilde{\boldsymbol{Q}}+b_1\tilde{\boldsymbol{Q}}^2\right)|\Delta(t)|^2 \\
     +& \left(\boldsymbol{a}_2\tilde{\boldsymbol{Q}}+b_2\tilde{\boldsymbol{Q}}^2\right)|\Delta(t)|^4,\label{eq:action}
\end{split}
\end{equation}
where $\tilde{\boldsymbol{Q}}=\boldsymbol{Q}+\boldsymbol{A}(t)$ is the gauge-invariant condensate momentum and  $\boldsymbol{Q}$ is the phase gradient of the order parameter.  $S_0$ is the zeroth order term in $\tilde{\boldsymbol{Q}}$, and $\Delta(t)$ the time dependent order parameter $\Delta(t)=\Delta_0+\delta\Delta(t)$.
 The constants $b_{1,2}$ are the usual Ginzburg-Landau coefficients appearing in even-power terms of $\tilde{\boldsymbol{Q}}$.  Linear-in-$\tilde{\boldsymbol{Q}}$ terms, $\boldsymbol{a}_1$ and $\boldsymbol{a}_2$, are only allowed in superconductors with broken time-reversal and inversion symmetries, and are  related to the Lifshitz invariant \cite{mineev2008nonuniform,bauer2012non}.
  
The action, Eq.~(\ref{eq:action}),  describes a helical superconductor with a spatially varying order parameter in the ground state, $\Delta(\boldsymbol{r})=\Delta_0 e^{i \boldsymbol{Q}_0 \boldsymbol{r}}$ \cite{agterberg2003novel,kaur2005helical,dimitrova2007theory}. The amplitude of modulation $\boldsymbol{Q}_0$ can be determined from the condition that the supercurrent $\boldsymbol{J}$ in the ground state must vanish: $\partial S/\partial \boldsymbol{Q}|_{\boldsymbol{Q}=\boldsymbol{Q}_0}=0$. Thus, $\boldsymbol{Q}_0=-(\boldsymbol{a}_1+\boldsymbol{a}_2 \Delta^2)/[2(b_1+b_2\Delta^2)]$. 
Next, we  calculate $\chi_{\boldsymbol{A}\Delta}$ by taking $\boldsymbol{Q}=\boldsymbol{Q}_0+\delta \boldsymbol{Q}$, where $\delta \boldsymbol{Q}$ is a phase gradient generated by passing a supercurrent through the system. Substituting Eq.~\eqref{eq:action} into Eq.~\eqref{eq:Higgslightcoupling} we obtain the Higgs-light coupling susceptibility
\begin{equation}
\chi_{\boldsymbol{A}\Delta}=\int d\boldsymbol{r} \left( 4 \Delta_0 \delta \boldsymbol{Q} (b_1+2b_2\Delta_0^2)+\frac{2(\boldsymbol{a}_2b_1-\boldsymbol{a}_1b_2)\Delta_0^3}{b_1+b_2\Delta_0^2} \right).
\label{eq:Chi}
\end{equation}
The first term describes the linear Higgs-light coupling due to a finite supercurrent discussed above and established in Ref.~\cite{moor2017amplitude}.  The second term is an additional contribution that is only finite in helical superconductors for which 
$\boldsymbol{a}_1$ and $\boldsymbol{a}_2$ are non-zero. This is one of the main results of our work: helical superconductors support linear Higgs-light coupling even in the absence of an applied supercurrent.  

Let us now investigate how the structure of the Higgs mode would change, if it were to be coupled linearly to an LC resonator (see setup in Fig.~\ref{Fig:Schematic}). The Higgs mode is described by the fluctuations of the order parameter $\delta\Delta(t)$, whereas the LC mode is described by the time-varying voltage $V(t)$. In frequency domain, the effective equations of motion of this system can be written as
  \begin{equation}
    \begin{pmatrix}
   \sqrt{\Omega_H-\Omega+i\Gamma_H} & \gamma_{1}\\
     \gamma_{2} & \Omega_{LC}-\Omega+i\Gamma_{LC}
    \end{pmatrix}
    \begin{pmatrix}
     \delta\Delta(\Omega) \\
     V(\Omega)
    \end{pmatrix}=0.\label{eq:coupling_matrix}
 \end{equation}
Here  $\Omega_H$ and $\Omega_{LC}$ are the resonant frequencies of the Higgs  and the LC mode, respectively, and $\Gamma_{H}$ and $\Gamma_{LC}$ are the damping parameters. The coupling coefficients  $\gamma_{1}$ and $\gamma_{2}$ are proportional to $\chi_{\boldsymbol{A}\Delta}$. We assume a vanishing injected DC supercurrent, which is why only the second term in Eq.~(\ref{eq:Chi}) contributes to the coupling. 

In the absence of any coupling, $\gamma_{1}=\gamma_{2}=0$, the Higgs mode is not a well-defined mode with a Lorentzian line-shape, as reflected in the square-root function in Eq.~\eqref{eq:coupling_matrix}. On the other hand, the LC mode is a regular mode.  
For a finite but small coupling such that ${\Gamma_H} \ll \text{Re}(\eta^2) \ll{\Omega_{LC}-\Omega_H}$, where $\eta=\frac{\gamma_{1 }\gamma_{2}}{\Omega_{LC}+i\Gamma_{LC}-\Omega_H}$, and a low dissipation of the $LC$ mode, $\Gamma_{LC}<\Omega_{LC}-\Omega_H$,  we can approximate the eigenvalue equation near the Higgs frequency as
\begin{equation}
    \sqrt{\Omega_H-\Omega+i\Gamma_H}-\eta=0.\label{eq:modified Higgs mode}
\end{equation}

If $\text{Re}(\eta)>0$, the eigenvalue equation can be linearized around the new resonance frequency and becomes
\begin{equation}
    \left[\Omega_H-\Re(\eta^2)-\Omega\right]  + i\left[\Gamma_H-\Im(\eta^2)\right]  = 0.
\end{equation}
The Higgs mode is shifted away from the branch cut at $\Omega_H$ to a lower frequency $\Omega_H-\Re(\eta^2)$, and becomes a real mode with a potentially small linewidth $\Gamma_H - \Im(\eta^2)$ \cite{linewidthnote}.
 Thus by coupling the two modes, the resonant response from the Higgs mode can be dramatically enhanced and its frequency reduced. This is our second main result. In what follows we derive our findings from a microscopic model.

\paragraph{Microscopic theory.-}One realization of  helical superconductivity is a quasi two-dimensional superconductor  with strong Rashba SOC and an in-plane magnetic field. To  calculate the susceptibility $\chi_{\boldsymbol{A}\Delta}$, we start with the generalized Eilenberger equation describing this system in the basis of the two helical bands labeled by the index $\lambda=\pm 1$ \cite{houzet2015quasiclassical}: 
\begin{multline}
i \boldsymbol{n}\cdot \left(\boldsymbol{Q}v_F/2+\lambda \boldsymbol{h}_\mathrm{ex} \times \hat{z}\right)[\tau_3,\hat{g}_{\lambda \boldsymbol{n}} ]
-\{\partial_t\tau_3,\hat{g}_{\lambda\boldsymbol{n}}\}
\\
= [\Delta_0 \tau_1+ \Delta_{\Omega}e^{-i\Omega t}\tau_1+ \hat{\Sigma}_{\lambda \boldsymbol{n}},\hat{g}_{\lambda \boldsymbol{n}}].
\label{eq:Eilenberger}
\end{multline}
Here, $\hat{g}_{\lambda\boldsymbol{n}}$ is the quasiclassical Green function of the band $\lambda$ with the momentum direction at the Fermi level given by $\boldsymbol{n}=\boldsymbol{p}/p_F$. It is a matrix in the Nambu space, spanned by the Pauli matrices $\tau_{1,2,3}$ and the identity matrix 1. The normalization condition $\hat{g}_{\lambda\boldsymbol{n}}^2=1$ holds. $\boldsymbol{Q}$ is the phase gradient of the order parameter, $\boldsymbol{h}_\mathrm{ex}$ is the exchange field, and $\hat{z}$ is the unit vector perpendicular to the plane of the superconductor. $v_F$ is the Fermi velocity, $\Delta_0$ is the saddle point value of the order parameter and  $\Delta_{\Omega}$ is the Fourier amplitude of the external pairing potential with frequency $\Omega$.
$\hat{\Sigma}_{\lambda \boldsymbol{n}}$ is the disorder self-energy  
\begin{equation}
 \hat{\Sigma}_{\lambda \boldsymbol{n}}=\sum_{\lambda^\prime}\frac{1}{4\tau_{\lambda^\prime}}\left(\langle\hat{g}_{\lambda^\prime\boldsymbol{n}^\prime}\rangle_{\boldsymbol{n}^\prime}+\lambda\lambda^\prime \boldsymbol{n} \cdot\langle\boldsymbol{n}^\prime\hat{g}_{\lambda^\prime\boldsymbol{n}^\prime}\rangle_{\boldsymbol{n}^\prime}\right).
\end{equation}
Here $\tau_{\lambda}$ is the effective scattering time for the band $\lambda$, $\frac{1}{\tau_{\lambda}}=\frac{1}{\tau}(1+\lambda\frac{\alpha}{v_F})$, where $\tau$ is the impurity scattering time. $\alpha$ is the velocity associated with Rashba SOC. Note that Eq.~\eqref{eq:Eilenberger} is valid when SOC is larger than all other energy scales except the chemical potential $\mu$, namely $\mu \gtrsim \alpha p_F\gg \Delta, h_{ex}, \tau^{-1}, Q v_F.$ In the absence of disorder, the two bands are decoupled, but share the same order parameter. Disorder couples the bands by introducing interband scattering.

\begin{figure}[h!]
\centering
\includegraphics[width = \columnwidth]{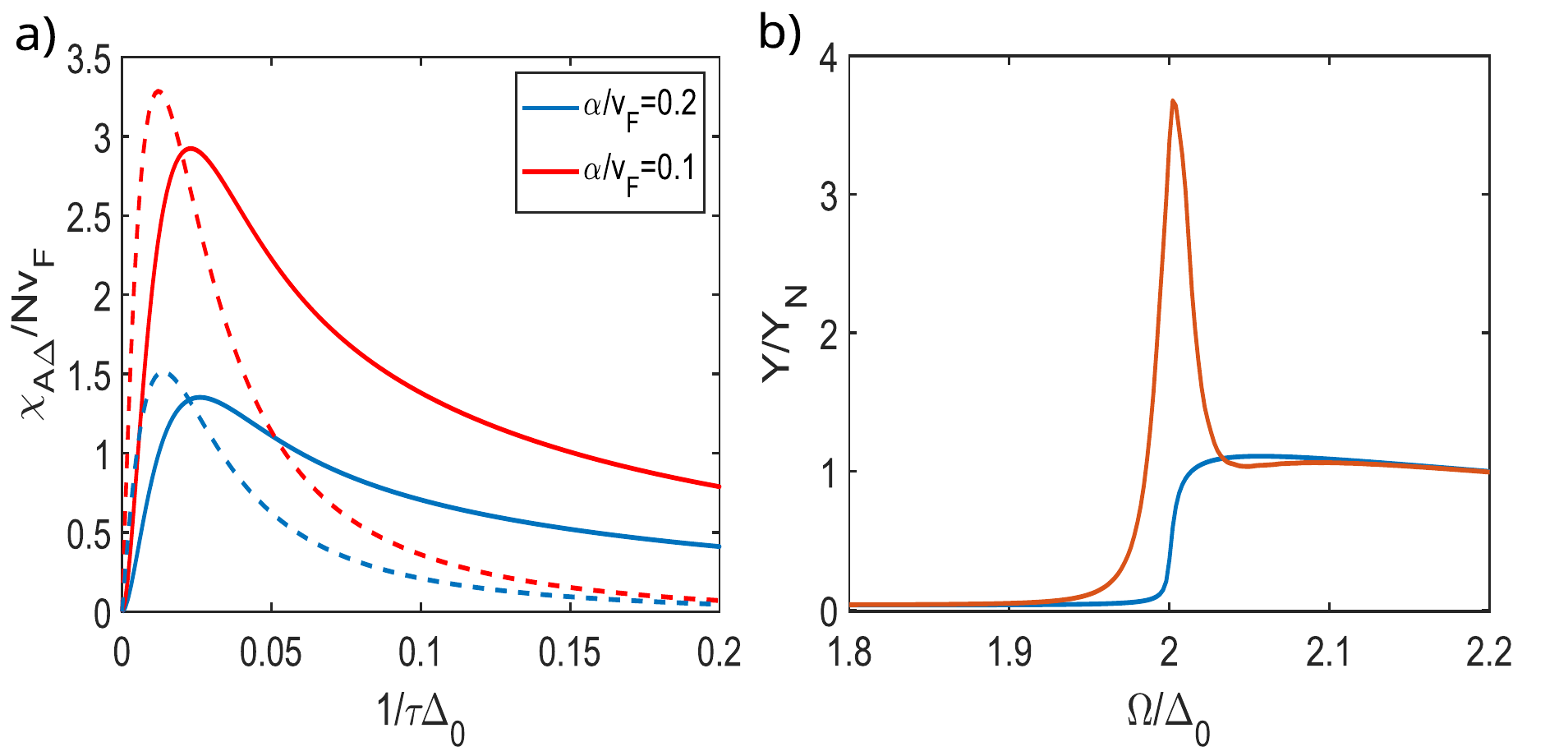}
\caption{a) Real part (solid lines) and imaginary part (dashed lines) of $\chi_{A\Delta}$ at $\Omega=2\Delta_0$ as a function of the disorder strength for different strengths of spin-orbit coupling.  $\chi_{A\Delta}$ behaves non-monotonically as a function of the disorder strength. b) Admittance normalized by its value in the normal state with (red line) and without (blue line) the Higgs contribution. The peak localized at $\Omega=2\Delta_0$ is the signature of the Higgs mode. The parameters used here are: $T=0$,  $h_\mathrm{ex}/\Delta_0=0.2$. In panel b), $1/\tau\Delta_0=0.02$, $\alpha/v_F=0.2$.}\label{Fig:chideA}
\end{figure}

Within linear response to the external field, the Green function can be written as $\hat{g}=\hat{g}^{(0)}e^{i\omega(t_1-t_2)}+\hat{g}^{(\Delta)}e^{i\omega_1t_1-i\omega_2t_2}$, with $\omega_1=\omega$ and $\omega_2=\omega+\Omega$. Here $\hat{g}^{(0)}$ is the unperturbative static Green function and $\hat{g}^{(\Delta)}$ denotes the external field-induced dynamical Green function, which is of the first order in $\Delta_{\Omega}$. Once we solve the Eilenberger equation, the supercurrent can be determined from

\begin{equation}
\boldsymbol{J}^{(0,\Delta)}=\sum_{\lambda}N_{\lambda} v_F \text{Tr}[\tau_3\langle\boldsymbol{n}\hat{g}^{(0,\Delta)}_{\lambda\boldsymbol{n}}\rangle_{\boldsymbol{n}}]. 
\end{equation}
Here, $\boldsymbol{J}^{(0)}$ and $\boldsymbol{J}^{(\Delta)}$ correspond to DC and AC supercurrents, respectively. $N_{\lambda}=N_0(1+\lambda\frac{\alpha}{v_F})$ is the density of states of band $\lambda$,  with $N_0$ denoting the average density of states. From the condition $\boldsymbol{J}^{(0)}(\boldsymbol{Q}=\boldsymbol{Q}_0)=0$, we first determine the modulation vector of the helical superconductivity $\boldsymbol{Q}_0$. Finally, we calculate $\chi_{\boldsymbol{A}\Delta}$ from
\begin{equation}
\chi_{\boldsymbol{A}\Delta}=\boldsymbol{J}^{(\Delta)}/\Delta_\Omega.
\label{eq:chi_AD}
\end{equation}

First we consider $\chi_{\boldsymbol{A}\Delta}$ in two limiting cases, namely, the pure ballistic  and the diffusive limits. In the ballistic limit $\tau^{-1}\to 0$, the system preserves a Galilean symmetry \cite{crowley2022supercurrent,papaj2022current}, so that the current is time-independent despite the presence of an external pairing field, and hence $\chi_{\boldsymbol{A}\Delta}=0$. On the other hand, in the diffusive limit $\tau^{-1}\gg \Delta, T$, the two helical bands are strongly mixed by disorder, and both bands are described by the same Usadel equation derived in Ref.~\cite{houzet2015quasiclassical}. This leads to a a suppression of $\chi_{\boldsymbol{A}\Delta}=0$  ( see  the supplementary material  \cite{supplementary})

For intermediate degree of disorder one needs to solve the Eilenberger equation, Eq. (\ref{eq:Eilenberger}).  The  weak exchange field allows for  an analytic solution presented in the supplement \cite{supplementary,code}. We find a non-monotonic behavior of $\chi_{\boldsymbol{A}\Delta}$ while increasing the disorder strength (Fig.~\ref{Fig:chideA}a). Without the disorder potential, $\chi_{\boldsymbol{A}\Delta}$ is zero, consistent with the above symmetry analysis. With increasing disorder strength, $\chi_{\boldsymbol{A}\Delta}$ rapidly reaches its maximum and then decays as a power law. We also verified that  $\chi_{\boldsymbol{A}\Delta}$ vanishes in the diffusive limit,  when $1/\tau\Delta_0\gg1$.

The linear Higgs-light coupling leads to a modification of the admittance of the superconductor. To find the total admittance, we write the order parameter as $\Delta=\Delta_0+\delta\Delta(t)$ and expand the action up to the second order in $\delta\Delta$ and the external field $\boldsymbol{A}$, $S=S_M+S_f$, where $S_M$ is the mean-field term and $S_f$ is the fluctuation term given by
\begin{equation}   S_f=\frac 1 T\sum_{n}
\left[\begin{array}{cc}
\delta\Delta(-\Omega_n) \\ \boldsymbol{A}(-\Omega_n)\end{array}\right]^\intercal
\left[\begin{array}{cc}
\mkern-5mu-\chi_{\Delta\Delta}^{-1} & \chi_{\Delta \boldsymbol{A}}\\
\chi_{\boldsymbol{A}\Delta} & \chi_{\boldsymbol{A}\boldsymbol{A}}
\end{array}\right]\left[\begin{array}{c}
\delta\Delta(\Omega_n)\\
\boldsymbol{A}(\Omega_n)
\end{array}\right],\label{eq:action_expansion}
\end{equation}
where $\Omega_n$ is the bosonic Matsubara frequency $\Omega_n=2n\pi T$ with $n\in \mathbb{Z}$. $\chi_{\Delta \boldsymbol{A}}$ is defined in Eq. (\ref{eq:chi_AD}), whereas 
$\chi_{\Delta\Delta}$ and $\chi_{\Delta \boldsymbol{A}}$ are defined as $\chi_{\Delta \boldsymbol{A}}=\partial F/\partial\boldsymbol{A}$, and $\chi_{\Delta\Delta}={\partial F}/{\partial \Delta_\Omega}$, where $F$ is the pair correlation $F=\sum_{\lambda}N_{\lambda}\text{Tr}[\tau_1\langle\hat{g}_{\lambda\boldsymbol{n}}\rangle_{\boldsymbol{n}}]$ . $\chi_{\Delta\boldsymbol{A}}$ and $\chi_{\boldsymbol{A}\Delta}$ are related by $\chi_{\Delta A}(\Omega)=\chi_{\boldsymbol{A}\Delta}(-\Omega)^*=\chi_{\boldsymbol{A}\Delta}(\Omega)$.
Finally, the field susceptibility is defined as $\chi_{\boldsymbol{A}\boldsymbol{A}}={\partial\boldsymbol{J}}/{\partial\boldsymbol{A}}$.

The pair susceptibility $\chi_{\Delta\Delta}$ has a square root singularity at $\Omega=2\Delta_0$ indicating the existence of the Higgs mode.  Integrating out the $\delta\Delta$ field, we obtain the total susceptibility 
\begin{equation}
    \tilde{\chi}_{\boldsymbol{A}\boldsymbol{A}}=\chi_{\boldsymbol{A}\boldsymbol{A}}+\chi_{\boldsymbol{A}\Delta}\chi_{\Delta\Delta}\chi_{\Delta \boldsymbol{A}}; ,  \label{eq:field susc}
\end{equation}
which defines the total admittance $Y=\tilde{\chi}_{\boldsymbol{A}\boldsymbol{A}}/i\Omega$. When $\chi_{\boldsymbol{A}\Delta}$ and $\chi_{\Delta \boldsymbol{A}}$ are finite, the admittance exhibits a peak at the Higgs frequency (Fig.~\ref{Fig:chideA}b)  providing a way of detecting the Higgs mode using standard experimental methods .

\begin{figure}[h!]
\centering
\includegraphics[width = 1\columnwidth]{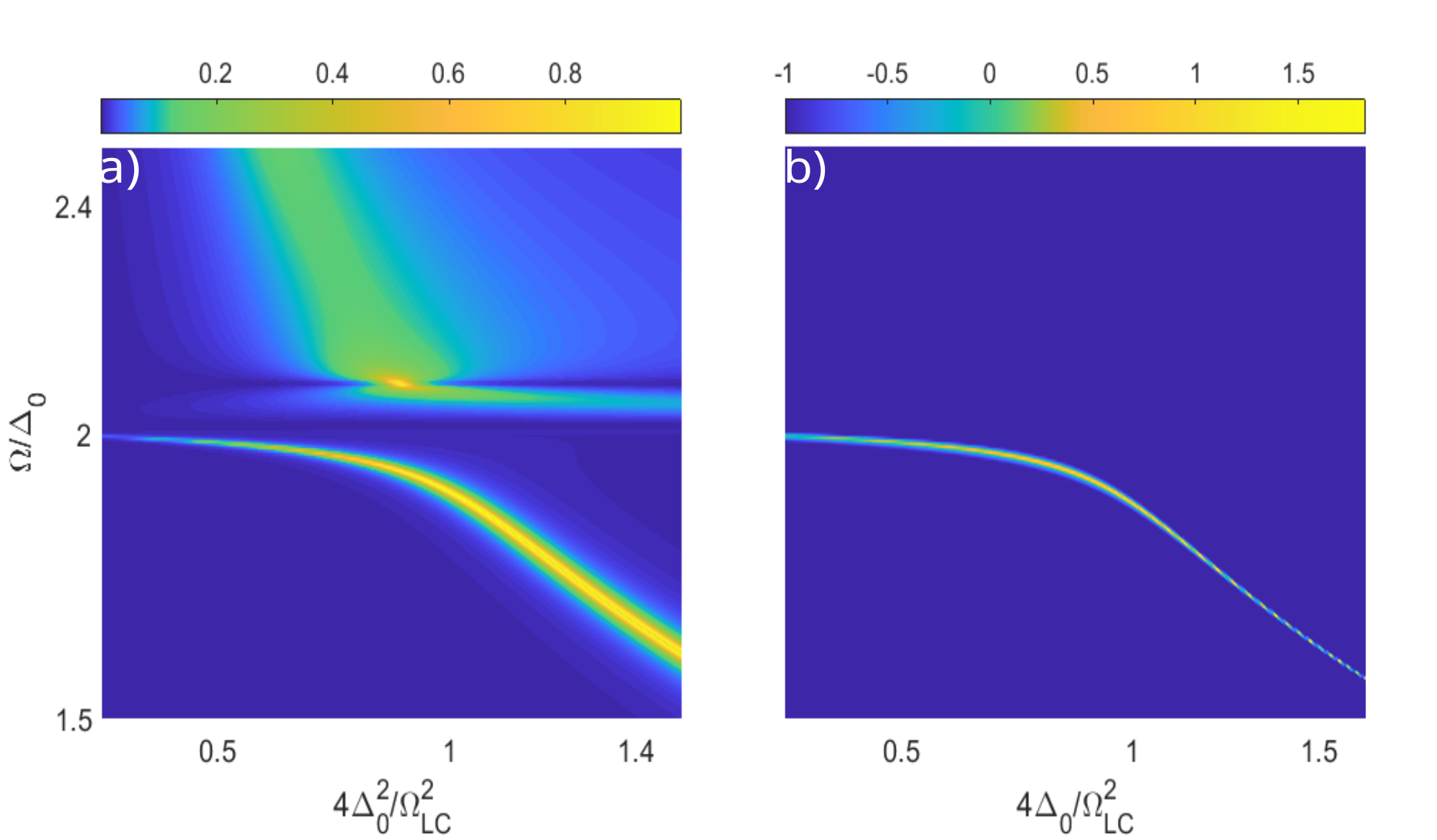}
\caption{a) The microwave absorption rate $W$ and b) the modified pair susceptibility $\tilde{\chi}_{\Delta\Delta}$ as functions of the inductance $L$ and the frequency $\Omega$.  Here we have assumed that the resistance is dominated by the resistance of the superconductor $R=R_S$. The parameters used here are: $T=0$, $\frac{1}{\tau\Delta_0}=0.1$, $\frac{h_\mathrm{ex}}{\Delta_0}=0.3$, $\frac{\alpha}{v_F}=0.3$,  $C=\frac{E_F}{\Delta_0^2}\frac{e_0^2}{2\pi}$,  $Z_t=100\frac{\Delta_0}{E_F}\frac{h}{e_0^2}$, where $E_F$ is the Fermi energy of the superconductor. Here we restore the electron charge $e_0$ and the Planck constant $h$ for clarity.}\label{Fig:spectrum}
\end{figure}

To couple the Higgs mode with an LC resonator we consider the circuit shown in Fig.~\ref{Fig:Schematic}. A capacitor and an inductor form an LC resonator. The total inductance of the circuit, $L^{-1}=L_{c}^{-1}+L_S^{-1}$, includes the inductance $L_c$ of the LC resonator and the kinetic inductance of the superconductor $L_S$. The total resistance is $R^{-1}=R_{c}^{-1}+R_S^{-1}$, where $R_{c}$ represents the damping of the LC circuit and $R_S$ is the resistance of the superconductor given by $R_S=i\Omega/\chi_{\boldsymbol{A}\boldsymbol{A}}$. We propose an experiment in which microwaves are sent to the system, for example through a transmission line, whereas the complex reflection coefficient is measured. To explicitly calculate the modified Higgs spectrum, we combine the equation of current  conservation  $I_C+I_{R}+I_L+I_{S}=I_\mathrm{ext}$, and the self-consistency equation for the dynamical part of the order parameter  
\begin{equation}
    \hat{M}\begin{pmatrix}
 \delta\Delta(\Omega)\\
 V(\Omega)/id\Omega 
\end{pmatrix}=\begin{pmatrix}
 0\\
 -I_\mathrm{ext}/Cd
\end{pmatrix}\, , \label{eq:equation of motion1}
\end{equation}
with the  response matrix given by 
\begin{equation}
    \hat{M}=\begin{pmatrix}
    \chi_{\Delta\Delta}^{-1} && -\chi_{\boldsymbol{A}\Delta} \\ 
    -\chi_{\Delta \boldsymbol{A}}\Omega_{LC}Z_0         && \Omega^2 -\Omega_{LC}^2-i\Omega\kappa ,
\end{pmatrix},\label{eq:equation of motion}
\end{equation}
where $\Omega_{LC}=\sqrt{1/LC}$, $Z_0=\sqrt{L/C}$ and $\kappa=1/RC$. The analytical expression of $\chi_{\Delta\Delta}$ was obtained in  Ref.~\cite{littlewood1982amplitude}. Its  general form is complicated, but for  $\Omega\lesssim 2\Delta$, $\chi_{\Delta\Delta}^{-1}$ scales as $\sqrt{2\Delta_0-\Omega}$. The system can thus be effectively described by Eq.~\eqref{eq:coupling_matrix}. Moreover, when $\Omega<\Omega_H=2\Delta$, $\chi_{\boldsymbol{A}\Delta}=\chi_{\Delta\boldsymbol{A}}$ become real \cite{supplementary}. The resonance frequency is determined by $\det \hat{M}=0$. The total impedance of the system is given by
\begin{equation}
 Z=\left[i\Omega C+\frac{1}{i\Omega L_c}+\frac{1}{R_c}+\frac{\tilde{\chi}_{\boldsymbol{A}\boldsymbol{A}}}{i\Omega}\right]^{-1},
\end{equation}
which determines the microwave reflection rate $r=(Z-Z_t)/(Z+Z_t)$, where $Z_t$ is the impedance of the transmission line and $\tilde{\chi}_{\boldsymbol{A}\boldsymbol{A}}$ is defined in Eq.~\eqref{eq:field susc}. The real part of $Z$ has peaks located at frequencies where $\det(\hat{M})=0$ showing that the resonant modes can be detected by measuring the impedance or the microwave reflection rate.

The microwave absorption rate $W$, defined as $W=1-|r|^2$, is shown in Fig.~\ref{Fig:spectrum}a. $W$ is  hugely enhanced at the frequencies of the resonance modes. An avoided crossing occurs due to linear coupling when the LC frequency matches the Higgs frequency. We find that the low-frequency mode is a well-defined mode with a frequency below the quasiparticle continuum, whereas the high-frequency mode is ill-defined and decays into quasiparticle excitations, especially when $\Omega\gtrsim 2\Delta_0$. Fig. ~\ref{Fig:spectrum}b shows the modified pair susceptibility $\chi^m_{\Delta\Delta }=\left(\chi_{\Delta\Delta}^{-1}+\chi_{\boldsymbol{A}\Delta}\chi_{\boldsymbol{A}\boldsymbol{A}}\chi_{\Delta\boldsymbol{A}}\right)^{-1} $, obtained by eliminating $V$ from Eq.~\eqref{eq:equation of motion1}. The low-frequency mode has a significant Higgs-component, especially when $LC<1/{(2\Delta_0)^2}$ and the mode occurs below the quasiparticle continuum.

\begin{figure}[h!]
\centering
\includegraphics[width = 0.8\columnwidth]{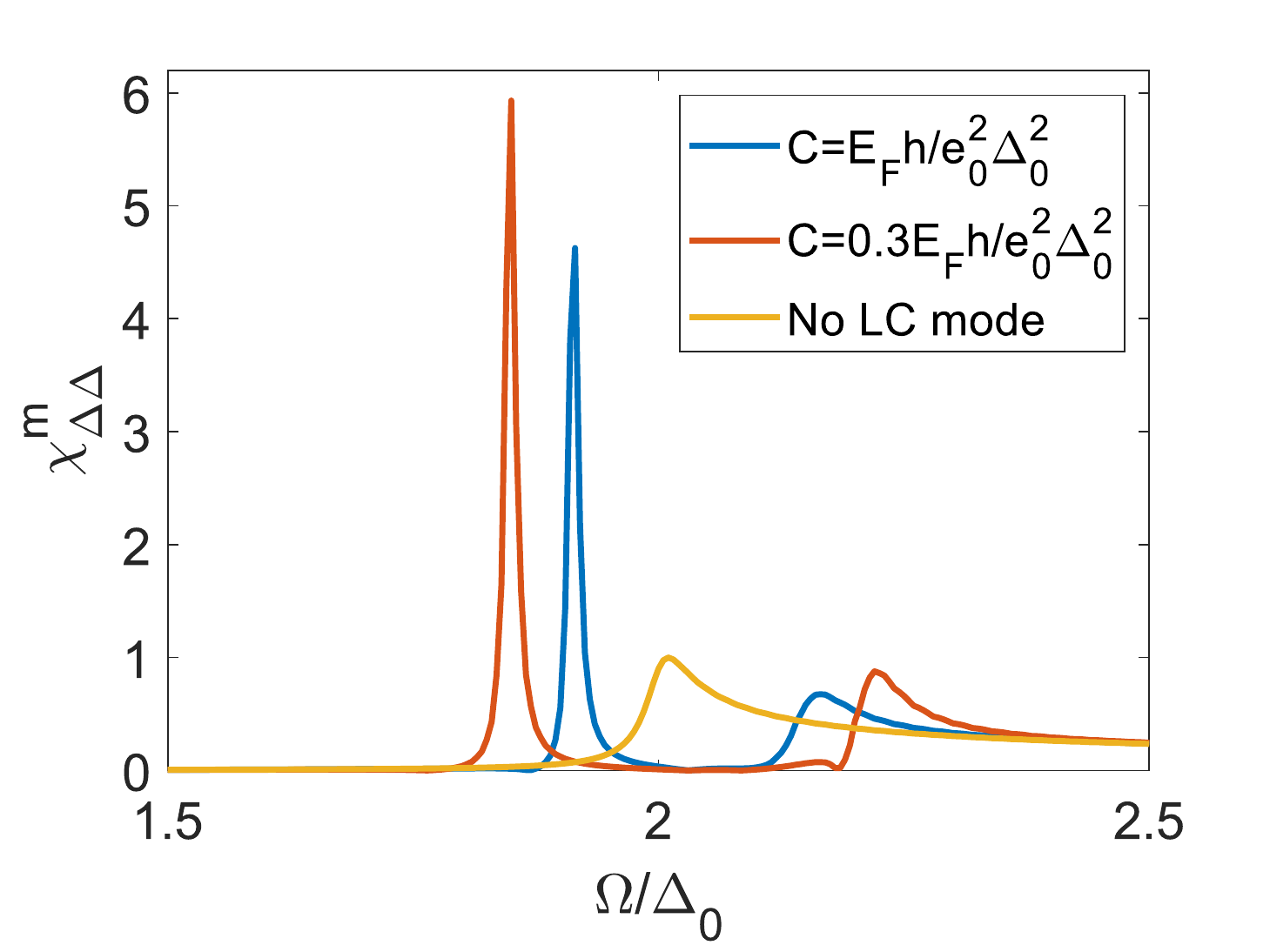}
\caption{The modified pair susceptibility $\chi_{\Delta\Delta}^m$ as a function of frequency for different values of $C$ normalized by the maximum value without the LC resonator.  The frequency of the LC resonator is fixed $\Omega_{LC}=\frac{1}{\sqrt{LC}}=2.1\Delta_0$. The parameters used here are the same as those in Fig.~\ref{Fig:spectrum}.}\label{Fig:differentC}
\end{figure}

Figure~\ref{Fig:differentC} shows how the spectral weight of the Higgs mode (pair susceptibility) depends on the capacitance $C$ of the LC circuit, with fixed LC frequency $\Omega_{LC}$. One can see how the pair susceptibility goes from a $\sqrt{2\Delta_0-\Omega}$ behaviour in the absence of the LC mode,  to 
a sharp resonance when coupled to the LC mode. The Higgs frequency is reduced with decreasing value of $C$. 

The suggested experiment  can be realized,  for example,  by galvanically coupling a  2D superconductor with strong SOC to a coplanar superconducting resonator \cite{wallraff2004strong}. The size of the latter  can be adjusted to be in resonance with the Higgs mode. Ideally, to avoid extra damping, the Higgs frequency $2\Delta_{HM}$ of the 2D superconductor needs to be smaller  than the  gap of  the superconductor forming the resonator.  Strong SOC can also be found at the LaAlO$_3$/SrTiO$_3$ interface. In this case
 $\Delta_{HM}\sim0.03$ meV \cite{reyren2009anisotropy}, which corresponds to a  bare Higgs mode frequency of around 10 GHz. This frequency  is accessible  with state-of-the art microwave measurement setups.  For the needed  Zeeman field, $\sim0.3\Delta_{HM}$ one can either apply an in-plane field  or  via magnetic proximity effect from an adjacent  ferromagnetic insulator like EuS\cite{wei2016strong,strambini2017revealing}.

\paragraph{Conclusion.-} We have shown that the linear Higgs-light coupling exists in a helical superconductor 
even without a supercurrent. From a phenomenological Ginzburg-Landau theory, we demonstrate that this linear Higgs-light coupling relies on the terms in the action related to the Lifshitz invariant. 
We confirm this result by explicitly calculating  the susceptibility $\chi_{\boldsymbol{A}\Delta}$ of a helical superconductor within a microsocopic theory. We find that $\chi_{\boldsymbol{A}\Delta}$ reaches its maximum at a weak disorder but vanishes in both clean and diffusive limits. We propose to reduce the mass of the Higgs mode by coupling it with an LC resonator. We demonstrate that the Higgs mode becomes an undamped regular collective mode when its frequency is reduced below the quasiparticle excitation energy $2\Delta_0$.

The linear Higgs-light coupling shows up in any system with a Lifshitz invariant. It may therefore be relevant also in the superconducting state of (twisted) multilayer graphene systems where the role of spin is replaced by the valley degree of freedom \cite{lin2022zero, xie2022valley}. On the other hand, it would be interesting to study this mechanism in multiband superconductors, where it might allow for a direct visualization of the amplitude modes.

\begin{acknowledgments}
      {\bf Acknowledgements}
      
Y.L., S.I., and F.S.B. acknowledge  financial support from Spanish AEI through project PID2020-114252GB-I00 (SPIRIT), and  the Basque Government through grant IT-1591-22 and IKUR strategy program.  F.S.B also acknowledges  the A. v. Humboldt Foundation.   R.O. was supported by the Deutsche Forschungsgemeinschaft (DFG, German Research Foundation) - TRR 288 - 422213477 (project A07). T.T.H. was supported by the Academy of Finland (Project No. 317118). The work of T. T. H , F. S. B., and S.I  is partially funded by the European Union’s Horizon research and innovation program under Grant Agreement No. 800923 (SUPERTED project). F.S.B. thanks Prof. Bj\"orn Trauzettel and his group for their kind hospitality during his stay in W\"urzburg University.

\end{acknowledgments}

\bibliography{refs}

\begin{thebibliography}{51}
\bibitem [{lin()}]{linewidthnote}%
  \BibitemOpen
  \href@noop {} {}\bibinfo {note} {In all models we have considered, ${\rm
  Im}(\eta^2)\le 0$ and therefore the overall linewidth stays
  positive.}\BibitemShut {Stop}%
\end{thebibliography}
\onecolumngrid
\pagebreak
\clearpage

\setcounter{equation}{0}
\setcounter{figure}{0}
\setcounter{table}{0}
\setcounter{page}{1}
\renewcommand{\theequation}{S\arabic{equation}}
\renewcommand{\thesection}{S\arabic{section}}
\renewcommand{\thefigure}{S\arabic{figure}}
\newcounter{defcounter}
    \newenvironment{defeq}{%
    \refstepcounter{defcounter}
    \renewcommand\theequation{S.\thedefcounter}
    \begin{equation}}
    {\end{equation}}

 \newenvironment{defalign}{%
  \refstepcounter{defcounter}
    \renewcommand\theequation{S.\thedefcounter}
    \begin{align}}
    {\end{align}}

 \newenvironment{defeqnarray}{%
  \refstepcounter{defcounter}
    \renewcommand\theequation{S.\thedefcounter}
    \begin{eqnarray}}
    {\end{eqnarray}}

\section{Supplementary Material}

In this supplementary material, we present the derivations of  the susceptibilities $\chi_{\boldsymbol A\Delta}$, and $\chi_{\boldsymbol A A}$, see Eq. (10) in the main text. Specifically we  present the solution of the Eilenberger equation within perturbation in $h$ and in $\boldsymbol{A}$.
We also show the frequency dependence of $\chi_{\boldsymbol{A}\Delta}$ and explain the vanishing of $\chi_{\boldsymbol{A}\Delta}$ in the diffusive limit.

\section{Calculation of $\chi_{\boldsymbol A\Delta}$}

We consider a helical superconductor realized in a 2D superconductor with large Rashba spin-orbit coupling (SOC) under an in-plane magnetic field. To  calculate the susceptibility $\chi_{\boldsymbol{A}\Delta}$, we start with the generalized Eilenberger equation describing this system in the basis of two helical bands labeled by the index $\lambda=\pm 1$ \cite{houzet2015quasiclassical}: 
\begin{defeq}
i \boldsymbol{n}\cdot \left(\boldsymbol{Q}v_F/2+\lambda \boldsymbol{h}_\mathrm{ex} \times \hat{z}\right)[\tau_3,\hat{g}_{\lambda \boldsymbol{n}} ]
-\{\partial_t\tau_3,\hat{g}_{\lambda\boldsymbol{n}}\}
= [\Delta_0 \tau_1+ \Delta_{\Omega}e^{-i\Omega t}\tau_1+ \hat{\Sigma}_{\lambda \boldsymbol{n}},\hat{g}_{\lambda \boldsymbol{n}}].
\label{eq:Eilenberger}
\end{defeq}
Here, $\hat{g}_{\lambda\boldsymbol{n}}$ is the quasiclassical Green function of the band $\lambda$ with the momentum direction at the Fermi level given by $\boldsymbol{n}=\boldsymbol{p}/p_F$. It is a matrix in the Nambu space, spanned by the Pauli matrices $\tau_{1,2,3}$ and the identity matrix 1. The normalization condition $\hat{g}_{\lambda\boldsymbol{n}}^2=1$ holds. $\boldsymbol{h}_\mathrm{ex}$ is the exchange field, and $\hat{z}$ is the unit vector perpendicular to the plane of the superconductor. $\boldsymbol{Q}$ is the phase gradient of the order parameter,
 $\boldsymbol{Q}=\boldsymbol{Q}_0+\delta\boldsymbol{Q}$, where $\boldsymbol{Q}_0$ is the anomalous phase gradient generated by $\boldsymbol{h}_{ex}$ and $\delta\boldsymbol{Q}$ is the supercurrent contribution. Here we assume zero DC supercurrent, so that $\boldsymbol{Q}=\boldsymbol{Q}_0$. $v_F$ is the Fermi velocity, $\Delta_0$ is the saddle point value of the order parameter,  and  $\Delta_{\Omega}$ is the Fourier transform of  the time-dependent  pairing potential driven by the external perturbation. The matrix $\hat{\Sigma}_{\lambda \boldsymbol{n}}$ is the self-energy describing the scattering off impurities.   
\begin{defeq}
 \hat{\Sigma}_{\lambda \boldsymbol{n}}=\sum_{\lambda^\prime}\frac{1}{4\tau_{\lambda^\prime}}\left(\langle\hat{g}_{\lambda^\prime\boldsymbol{n}^\prime}\rangle_{\boldsymbol{n}^\prime}+\lambda\lambda^\prime \boldsymbol{n} \cdot\langle\boldsymbol{n}^\prime\hat{g}_{\lambda^\prime\boldsymbol{n}^\prime}\rangle_{\boldsymbol{n}^\prime}\right).
\end{defeq}
Here $\tau_{\lambda}$ is the effective scattering time for the band $\lambda$, $\frac{1}{\tau_{\lambda}}=\frac{1}{\tau}(1+\lambda\frac{\alpha}{v_F})$, where $\tau$ is the impurity scattering time. $\alpha$ is  Rashba parameter describing the SOC. Note that Eq.~\eqref{eq:Eilenberger} is valid when SOC is larger than all other energy scales except the chemical potential $\mu$, namely $\mu \gtrsim \alpha p_F\gg \Delta, h_\mathrm{ex}, \tau^{-1}, Q v_F.$ In the absence of disorder, the two bands are decoupled, but share the same order parameter. Disorder couples the bands by introducing interband scattering. 

Up to  linear order in $\Delta_{\Omega}$, the Green function can be written as
\begin{defeq}
    \hat{g}_{\lambda\boldsymbol{n}}=\hat{g}_{\lambda\boldsymbol{n}}^{(0)}e^{i\omega(t_1-t_2)}+\hat{g}_{\lambda\boldsymbol{n}}^{(\Delta)}e^{i\omega_1t_1-\omega_2t_2},
\end{defeq}
with $\omega_1=\omega$ and $\omega_2=\omega+\Omega$. Here $\omega$ is the fermion Matsubara  frequency $\omega=(2n+1)\pi T$, where $n$ is an integer and $T$ is the temperature while $\Omega$ is the boson Matsubara frequency $\Omega=2n\pi T$. $\hat{g}^{(0)}$ is the static Green function and $\hat{g}^{(\Delta)}$ denotes the external field induced dynamical Green function, which is linear in $\Delta_{\Omega}$. Here we consider a small exchange field limit and treat it as a perturbation. The zeroth order solution in $\boldsymbol{h}_\mathrm{ex}$ can be easily obtained
\begin{defeq}
    \hat{g}_{\lambda\boldsymbol{n}}^{(0)}(t)=\hat{g}_{\lambda\boldsymbol{n}}^{(00)}e^{i\omega(t_1-t_2)}+\hat{g}_{\lambda\boldsymbol{n}}^{(\Delta 0)}e^{i\omega t_1-i\omega_2t_2},
\end{defeq}
with
%
\begin{defeq}
\begin{aligned}
    \hat{g}_{\lambda\boldsymbol{n}}^{(00)}(\omega)&=\frac{\omega\tau_3+\Delta\tau_1}{\sqrt{\omega^2+\Delta^2}},\\
   \hat{g}_{\lambda\boldsymbol{n}}^{(\Delta 0)}(\omega_1,\omega_2)&=\Delta_{\Omega}\frac{\hat{g}_{\lambda\boldsymbol{n}}^{(00)}(\omega_1)\tau_1\hat{g}_{\lambda\boldsymbol{n}}^{(00)}(\omega_2)-\tau_1}{s_1+s_2},
\end{aligned}\label{eq:othGF}
\end{defeq}

and
\begin{defeq}
\begin{aligned}
    s_1\equiv s(\omega_1)=\sqrt{\omega_1^2+\Delta^2},\\
    s_2\equiv s(\omega_2)=\sqrt{\omega_2^2+\Delta^2}.
\end{aligned}
\end{defeq}
Note that without the exchange field, the quasiclassical Green function is isotropic and the disorder self-energy has no effect on the Green functions. 
Next, we determine corrections to the static and dynamical  Green's functions, ${\boldsymbol{g}}_{\lambda \boldsymbol{n} }^{(0h)}$ and $\hat{\boldsymbol{g}}_{\lambda\boldsymbol{n}}^{(\Delta h)}$,   in  leading order with respect to  $h_{ex}$.

For  small $\boldsymbol{h}_{ex}$, the Green's function is almost isotropic, and we  can approximate the Green's function up to the first two terms of the 2D harmonic  expansion:
\begin{defeq}
\begin{aligned}
\hat{g}_{\lambda\boldsymbol{n}}^{(0)}&=\hat{g}^{(00)}+\hat{\boldsymbol{g}}_{\lambda }^{(0h)}\cdot\hat{\boldsymbol{n}},\\
\hat{g}_{\lambda\boldsymbol{n}}^{(\Delta)}&=\hat{g}^{(\Delta 0)}+\hat{\boldsymbol{g}}_{\lambda}^{(\Delta h)}\cdot\hat{\boldsymbol{n}},
\end{aligned}
\end{defeq}
where $\hat{g}^{(00)}$ and $\hat{g}^{(\Delta 0)}$ are the isotropic components of the Green functions defined in Eqs.(\ref{eq:othGF}), and we neglect here the indices $\lambda$ and $\boldsymbol{n}$.
We first focus on the static part of the Green's function,  $\hat{g}_{\lambda\boldsymbol{n}}^{(0)}$. Linearization of Eq. (\ref{eq:Eilenberger}) with respect to  $\boldsymbol{h_{ex}}$ gives
\begin{defeq} 
\begin{aligned}
2i\boldsymbol{v}_F\boldsymbol{Q}_+\frac{\Delta}{s}\tau_2&=s[\hat{g}^{(00)},\hat{\boldsymbol{g}}_+^{(0h)}]+\left(\frac{1}{4\tau_-}+\frac{1}{8\tau_+}\right)[\hat{g}^{(00)},\hat{\boldsymbol{g}}_+^{(0h)}]+\frac{1}{8\tau_-}[\hat{g}^{(00)},\hat{\boldsymbol{g}}_-^{(0h)}],\\
2i\boldsymbol{v}_F\boldsymbol{Q}_-\frac{\Delta}{s}\tau_2&=s[\hat{g}^{(00)},\hat{\boldsymbol{g}}_-^{(0h)}]+\left(\frac{1}{4\tau_+}+\frac{1}{8\tau_-}\right)[\hat{g}^{(00)},\hat{\boldsymbol{g}}_-^{(0h)}]
+\frac{1}{8\tau_+}[\hat{g}^{(00)},\hat{\boldsymbol{g}}_+^{(0h)}].
\end{aligned} \end{defeq}
For convenience, we have defined $\boldsymbol{Q}_h=\boldsymbol{h}_\mathrm{ex}\times\hat{z}/v_F$ and $\boldsymbol{Q}_{\pm}=\boldsymbol{Q}_0\pm \boldsymbol{Q}_h$. We have also used the fact that $Q_0$ is of the same order of $\boldsymbol{h}_{ex}$.
The solution of these equations reads
\begin{defeq} \begin{aligned}
 \hat{\boldsymbol{g}}_+^{(0h)}&=\frac{iv_F\Delta\hat{g}^{(00)}\tau_2\boldsymbol{X}_+}{sY},\\
\hat{\boldsymbol{g}}_-^{(0h)}&=\frac{iv_F\Delta\hat{g}^{(00)}\tau_2\boldsymbol{X}_-}{sY},
\end{aligned} \end{defeq}
with
\begin{defeq} \begin{aligned}
    \boldsymbol{X}_+&=\left[\left(s+\frac{1}{4\tau_+}+\frac{1}{8\tau_-}\right)\boldsymbol{Q}_+-\frac{1}{8\tau_-}\boldsymbol{Q}_-\right],\\
    \boldsymbol{X}_-&=\left[\left(s+\frac{1}{4\tau_-}+\frac{1}{8\tau_+}\right)\boldsymbol{Q}_--\frac{1}{8\tau_+}\boldsymbol{Q}_+\right],
\end{aligned} \end{defeq}
and
\begin{defeq}
    Y=\left(s+\frac{1}{4\tau_-}+\frac{1}{8\tau_+}\right)\left(s+\frac{1}{4\tau_+}+\frac{1}{8\tau_-}\right)-\frac{1}{8\tau_+}\frac{1}{8\tau_-}.
\end{defeq}
Here $s=\sqrt{\omega^2+\Delta^2}$. This static Green function is used to determine the anomalous phase gradient $\boldsymbol{Q}_0$ above Eq. (10) in the main text.


To get the dynamic part of the Green's function, we expand the Eilenberger equation. Eq. (\ref{eq:Eilenberger}),  in $\Delta_\Omega$ and $\boldsymbol{Q}$. Linear terms give:
\begin{defeqnarray}
    v\boldsymbol{Q}_+[\tau_3,\hat{{g}}_{+}^{(\Delta 0)}]&=&s[\hat{g}^{(00)}, \hat{\boldsymbol{g}}_{+}^{(\Delta h)}]
    +\left(\frac 1 {4\tau_-}+\frac{1}{8\tau+}\right)[\hat{g}^{(00)},\hat{\boldsymbol{g}}_{+}^{(\Delta h)}]
      +\left(\frac 1 {4\tau_-} + \frac{1}{8\tau+}\right)[\hat{g}^{(\Delta 0)},\hat{\boldsymbol{g}}_{+}^{(0h)}]\nonumber\\
        &+&\frac 1 {8\tau_-}[\hat{g}^{(00)},\hat{\boldsymbol{g}}_{-}^{(\Delta h)}]
      +\frac 1 {8\tau_-}[\hat{g}^{(\Delta 0)},\hat{\boldsymbol{g}}_{-}^{(0 h)}]+[\Delta_{\Omega}\tau_1,\hat{\boldsymbol{g}}_+^{(0h)}],
\end{defeqnarray}
and 
\begin{defeqnarray}
    v\boldsymbol{Q}_-[\tau_3,\hat{g}_{-}^{(\Delta 0)}]&=&s[\hat{g}^{(00)}, \hat{\boldsymbol{g}}_{-}^{(\Delta h)}]
    +\left(\frac 1 {4\tau_+}+\frac{1}{8\tau-}\right)[\hat{g}^{(00)},\hat{\boldsymbol{g}}_{-}^{(\Delta h)}]
      +\left(\frac 1 {4\tau_+}+\frac{1}{8\tau-}\right)[\hat{g}^{(\Delta 0)},\hat{\boldsymbol{g}}_{-}^{(0h)}]\nonumber\\
        &+&\frac 1 {8\tau_+}[\hat{g}^{(00)},\hat{\boldsymbol{g}}_{+}^{(\Delta h)}]
      +\frac 1 {8\tau_+}[\hat{g}^{(\Delta 0)},\hat{\boldsymbol{g}}_{+}^{(0h)}]
    +[\Delta_{\Omega}\tau_1,\hat{\boldsymbol{g}}_-^{(0h)}].
\end{defeqnarray}
The solution reads
\begin{defeqnarray}
  \hat{\boldsymbol{g}}_{+}^{(\Delta h)}=\frac{\hat{g}^{(00)}\left[\left(s_1+s_2+\frac{1}{2\tau_+}+\frac{1}{4\tau_-}\right)\boldsymbol Z_+-\frac{1}{4\tau_-}\boldsymbol Z_-\right]}{Y_2}\Delta_{\Omega},
\end{defeqnarray}

\begin{defeqnarray}
  \hat{\boldsymbol{g}}_{-}^{(\Delta h)}=\frac{\hat{g}^{(00)}\left[\left(s_1+s_2+\frac{1}{2\tau_-}+\frac{1}{4\tau_+}\right)\boldsymbol Z_--\frac{1}{4\tau_+}\boldsymbol Z_+\right]}{Y_2}\Delta_{\Omega},
\end{defeqnarray}

with
\begin{defeqnarray}
Y_2&=&\left(s_1+s_2+\frac 1 {2\tau_-} +\frac 1 {4\tau_+}\right)\left(s_1+s_2+\frac 1 {2\tau_+}+ \frac 1 {4\tau_-}\right)-\frac{1}{4\tau_+}\frac{1}{4\tau_-},
\end{defeqnarray}
\begin{defeqnarray}
\boldsymbol{Z}_+&=&v\boldsymbol{Q}_+[\tau_3,\hat{g}_{+}^{(\Delta 0)}]-\left(\frac 1 {4\tau_-} + \frac 1 {8\tau_+}\right)[\hat{g}^{(\Delta 0)},\hat{\boldsymbol{g}}_+^{(0h)}]
-\frac 1 {8\tau_-}[\hat{g}^{(\Delta 0)},\hat{\boldsymbol{g}}_-^{(0h)}]+[\Delta_{\Omega}\tau_1e^{i\Omega t},\hat{\boldsymbol{g}}_+^{(0h)}]\nonumber\\
&-&\left(s_2+\frac 1 {4\tau_-}+\frac{1}{8\tau_+}\right)\{\hat{g}^{(\Delta 0)},\hat{\boldsymbol{g}}_{+}^{(0h)}\}
-\frac 1 {8\tau_-}\{\hat{g}^{(\Delta 0)},\hat{\boldsymbol{g}}_-^{(0h)}\},\label{eq:Zplus}
\end{defeqnarray}
and
\begin{defeqnarray}
\boldsymbol{Z}_-&=&v\boldsymbol{Q}_-[\tau_3,\hat{g}_{-}^{(\Delta 0)}]-\left(\frac 1 {4\tau_+}+ \frac 1 {8\tau_-}\right)[\hat{g}^{(\Delta 0)},\hat{\boldsymbol{g}}_-^{(0h)}]
-\frac 1 {8\tau_+}[\hat{g}^{(\Delta 0)},\hat{\boldsymbol{g}}_+^{(0h)}]+[\Delta_{\Omega}\tau_1e^{i\Omega t},\hat{\boldsymbol{g}}_-^{(0h)}]\nonumber\\
&-&\left(s_2+\frac 1 {4\tau_+}+\frac{1}{8\tau_-}\right)\{\hat{g}^{(\Delta 0)},\hat{\boldsymbol{g}}_{-}^{(0h)}\}
-\frac 1 {8\tau_+}\{\hat{g}^{(\Delta 0)},\hat{\boldsymbol{g}}_+^{(0h)}\}.\label{eq:Zminus}
\end{defeqnarray}
The pair-spin susceptibility is then given by

\begin{defeq}
\chi_{\boldsymbol{A}\Delta}=\sum_{\lambda}N_{\lambda}\text{Tr}\left[v_F\tau_3\hat{\boldsymbol{g}}_{\lambda}^{(\Delta h)}\right]/\Delta_\Omega.
\label{deq:chi}
\end{defeq}

The spin-pair susceptibility $\chi_{\Delta\boldsymbol{A}}$ is related to $\chi_{\boldsymbol{A}\Delta}$ by $\chi_{\Delta\boldsymbol{A}}(\Omega)=\chi_{\boldsymbol{A}\Delta}(-\Omega)^*=\chi_{\boldsymbol{A}\Delta}(\Omega)$. These expressions of $\chi_{\boldsymbol{A}\Delta}$ and $\chi_{\Delta\boldsymbol{A}}$ are used in plotting Fig.~(2-4) in the main text.

\section{calculation of $\chi_{\boldsymbol A\boldsymbol A}$}

In order to find $\chi_{\boldsymbol A\boldsymbol A}$ we need to consider the Eilenberger equation with an external electromagnetic field $\boldsymbol A_{\Omega}$ 
\begin{defeq}
i \boldsymbol{n}\cdot \left(\boldsymbol{Q}_0v_F/2+\lambda \boldsymbol{h}_\mathrm{ex} \times \hat{z}\right)[\tau_3,\hat{g}_{\lambda \boldsymbol{n}} ]
-\{\partial_t\tau_3,\hat{g}_{\lambda\boldsymbol{n}}\}
= [\Delta_0 \tau_1+ \boldsymbol A_{\Omega}\cdot\boldsymbol{n}\tau_3e^{-i\Omega t}+ \hat{\Sigma}_{\lambda \boldsymbol{n}},\hat{g}_{\lambda \boldsymbol{n}}].
\label{eq:Eilenberger2}
\end{defeq}
Unlike calculating $\chi_{\boldsymbol A\Delta}$, here we only consider the zeroth order term in $\boldsymbol{h}_\mathrm{ex}$.  We expand the Green function as
\begin{defeq}
    \hat{g}_{\lambda\boldsymbol{n}}=\hat{g}^{(0)}e^{i\omega(t_1-t_2)}+\boldsymbol{n}\cdot\hat{\boldsymbol g}_{\lambda}^{(A)}e^{i\omega_1t_1-i\omega_2t_2}.
\end{defeq}
where $\hat{g}^{(0)}$ is the static Green function and $\hat{\boldsymbol g}_{\lambda}^{(A)}$ is the dynamical Green function induced by $\boldsymbol{A}_{\Omega}$.
Keeping only the first order terms in $\boldsymbol{A}_{\Omega}$ we have

\begin{defeqnarray}
   s_1\hat{g}^{(0)}\hat{\boldsymbol g}_+^{(A)}-s_2\hat{\boldsymbol g}_{+}^{(A)}\hat{g}^{(0)}=\boldsymbol A_\Omega[\tau_3,\hat{g}^{(0)}]-\left(\frac{1}{4\tau_-}+\frac{1}{8\tau_+}\right)[\hat{g}^{(0)},\hat{\boldsymbol g}_{+}^{(A)}]
-\frac{1}{8\tau_-}[\hat{g}^{(0)},\hat{\boldsymbol g}_{-}^{(A)}],\\
   s_1\hat{g}^{(0)}\hat{\boldsymbol g}_{-}^{(A)}-s_2\hat{\boldsymbol g}_{-}^{(A)}\hat{g}^{(0)}=\boldsymbol A_\Omega[\tau_3,\hat{g}^{(0)}]-\left(\frac{1}{4\tau_+}+\frac{1}{8\tau_-}\right)[\hat{g}^{(0)},\hat{\boldsymbol g}_{-}^{(A)}]
-\frac{1}{8\tau_+}[\hat{g}^{(0)},\hat{\boldsymbol g}_{+}^{(A)}].
\end{defeqnarray}

Solving these equations, we have
\begin{defeqnarray}
   \hat{\boldsymbol g}_{+}^{(A)}=\boldsymbol A_\Omega\frac{M_+\left[\hat{g}^{(0)}(\omega_1)\tau_3\hat{g}^{(0)}(\omega_2)-\tau_3\right]}{N_+},\\
   \hat{\boldsymbol g}_{-}^{(A)}=\boldsymbol A_\Omega\frac{M_-\left[\hat{g}^{(0)}(\omega_1)\tau_3\hat{g}^{(0)}(\omega_2)-\tau_3\right]}{N_-},
\end{defeqnarray}
with
\begin{defeq} \begin{aligned}
    M_\pm&=s_1+s_2+\frac{1}{4\tau_\pm},\\
    N_+&=\left(s_1+s_2+\frac 1 {4\tau_-}+\frac 1 {8\tau_\pm}\right)\left(s_1+s_2+\frac 1 {4\tau_\pm}+\frac 1 {8\tau_\mp}\right)-\frac 1 {64\tau_\pm\tau_\mp}.
\end{aligned} \end{defeq}
The field susceptibility $\chi_{\boldsymbol{A}\boldsymbol{A}}$ is given by 
\begin{defeq}
    \chi_{\boldsymbol{A}\boldsymbol{A}}=\sum_{\lambda}N_{\lambda}\text{Tr}\left[v_F\tau_3\hat{g}_\lambda^{(A)}\right]/A_\Omega.
\end{defeq}
This expression of $\chi_{\boldsymbol{A}\boldsymbol{A}}$ is used in plotting Fig. (2-3) in the main text.

\section{frequency dependence of $\chi_{\boldsymbol{A}\Delta}$}
\begin{figure}[h!]
\centering
\includegraphics[width = 0.6\columnwidth]{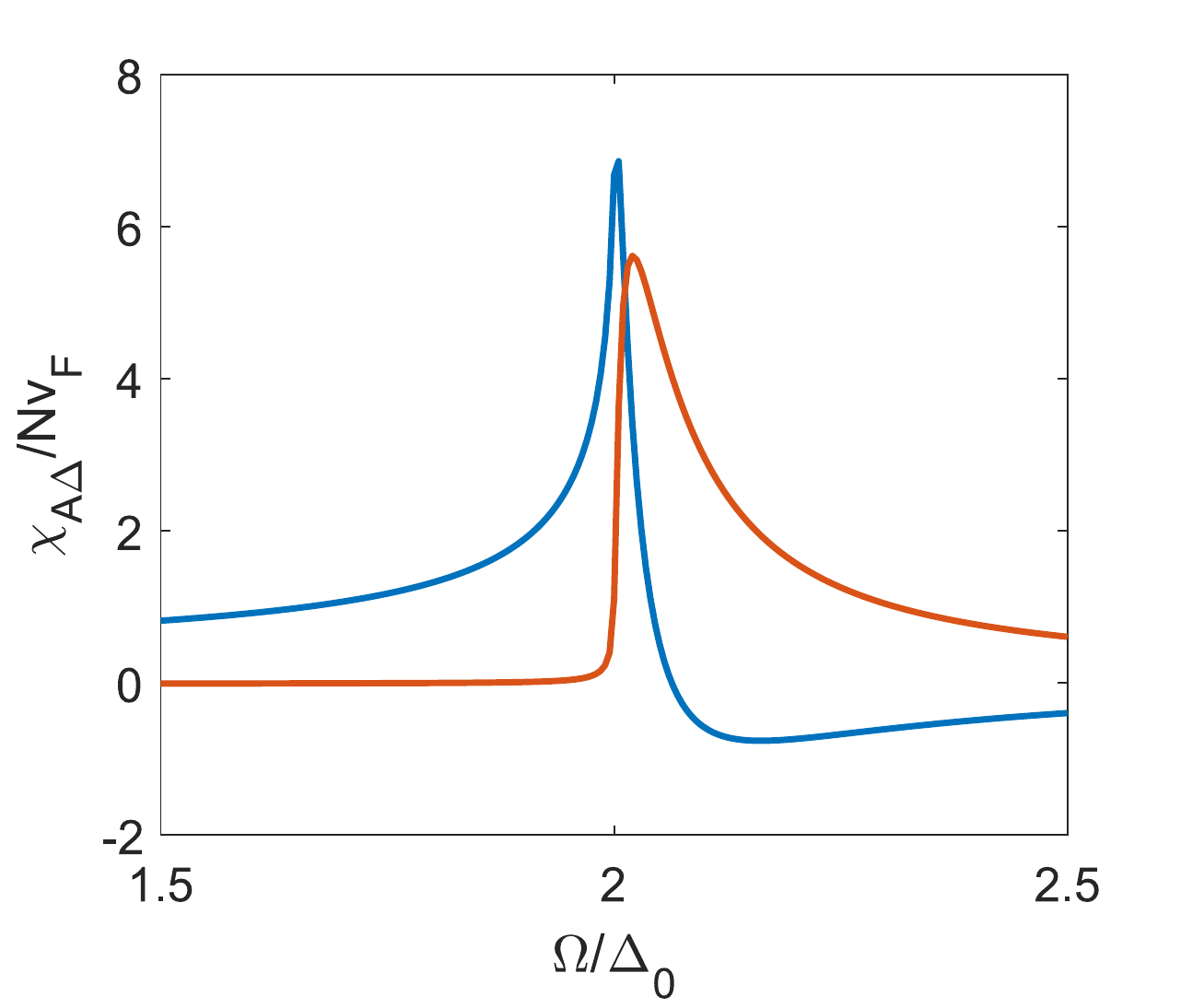}
\caption{The pair spin susceptibility $\chi_{\boldsymbol{A}\Delta}$ as a function of frequency. The blue and red lines denotes the real and imaginary parts, respectively. The parameters used here are:  $T=0$,  $h_\mathrm{ex}/\Delta_0=0.3$.  $1/\tau\Delta_0=0.1$, $\alpha/v_F=0.3$.} \label{Fig:frequency dependence}
\end{figure}

We have shown in the main text that in helical superconductors both the real part and imaginary part of $\chi_{\boldsymbol{A}\Delta}$ can be finite at $\Omega=2\Delta_0$. Here, we calculate the full frequency dependence of $\chi_{\boldsymbol{A}\Delta}$, obtained from Eq. (\ref{deq:chi}). The results are shown in Fig.~\ref{Fig:frequency dependence}.

One can  see that $\chi_{\boldsymbol{A}\Delta}$ is  real when $\Omega<2\Delta_0$. On the other hand, at high frequencies $\Omega>2\Delta_0$, the imaginary part of $\chi_{\boldsymbol{A}\Delta}$ dominates.

\section{Vanishing of $\chi_{\boldsymbol A\Delta}$ in the diffusive limit}

In the diffusive limit $\alpha p_F \gg \tau^{-1} \gg \Delta_0,T, h_{ex}$, the two helical bands are strongly mixed by disorder.  Both helical bands are described by the same quasiclassical Green's function averaged over the Fermi surface, $\hat{g}=\langle \hat{g}_{\pm1 \boldsymbol{n}}\rangle_{\boldsymbol{n}}$, which  satisfies the Usadel equation \cite{houzet2015quasiclassical}
\begin{equation}
-\frac{D}{4} [\boldsymbol{Q}'\tau_3, \hat{g} [\boldsymbol{Q}'\tau_3,\hat{g}]]
-\{\partial_t\tau_3,\hat{g}\}
= [\Delta_0 \tau_1+ \Delta_{\Omega}e^{-i\Omega t}\tau_1+\Gamma \tau_3 \hat{g} \tau_3,\hat{g}].
\label{eq:Usadel}
\end{equation}
Here, the role of the magnetic field is two-fold. First, it introduces the a shift to the gauge-invariant condensate momentum $\boldsymbol{Q}'=\boldsymbol{Q}+\boldsymbol{A}_\mathrm{eff}$, where $\boldsymbol{A}_\mathrm{eff}=4\alpha\boldsymbol{h}_\mathrm{ex}\times\hat{z}/(\alpha^2+v_F^2)$. Second, it introduces a depairing rate  $\Gamma=2 \tau h_\mathrm{ex}^2(v_F^2-\alpha^2)/(v_F^2+\alpha^2) $. The diffusion constant is $D=\tau(v_F^2+\alpha^2)/2$.

In the ground state, the minimization of the energy requires that the gauge-invariant condensate momentum vanishes, and therefore $\boldsymbol{Q}_0+\boldsymbol{A}_\mathrm{eff}=0$. From here, we see that the only effect of the magnetic field is to introduce depairing rate $\Gamma$, and Higgs-light coupling is absent: $\chi_{\boldsymbol{A}\Delta}=0$.


\end{document}